# The Case for a Solar Influence on Certain Nuclear Decay Rates

P.A. Sturrock[1], E. Fischbach[2], D. Javorsek II[3], J.H. Jenkins[2,4], R.H. Lee[5]
[1] Center for Space Science and Astrophysics, Stanford University, Stanford, CA 94305-4060, USA
[2] Department of Physics, Purdue University, West Lafayette, IN 47907, USA
[3] Edwards Air Force Base, CA 93524, USA
[4] School of Nuclear Engineering, Purdue University, West Lafayette, IN 47907
[5] Department of Physics, United States Air Force Academy, Colorado Springs, CO 80920

Power-spectrum analyses of the decay rates of certain nuclides reveal (at very high confidence levels) an annual oscillation and periodicities that may be attributed to solar rotation and to solar r-mode oscillations. A comparison of spectrograms (time-frequency displays) formed from decay data and from solar neutrino data reveals a common periodicity with frequency 12.5 year$^{-1}$, which is indicative of the solar radiative zone. We propose that the neutrino flux is modulated by the solar magnetic field (via Resonant Spin Flavor Precession) in that region, and we estimate the force and the torque that could be exerted on a nuclide by the solar neutrino flux.

## 1 . Introduction

Jenkins *et al.*[1], Fischbach *et al.*[2], and Javorsek *et al.*[3] have analyzed data from measurements of the decay rates of $^{32}$Si and $^{36}$Cl acquired at the Brookhaven National Laboratory (BNL) [4] and from measurements of the decay rate of $^{226}$Ra acquired at the Physikalisch-Techniche Bundesanstalt (PTB) [5], finding evidence of small but statistically significant annual variations in those rates. These findings led those authors to suggest that nuclear decay rates may be influenced (via some unknown mechanism) by the Sun. The evidence for a solar influence has been strengthened by the discovery of the influence of solar rotation[6,7] and of internal solar r-mode oscillations[8], to be discussed in Section 2. These results are suggestive of the influence of solar neutrinos. As a check of this hypothesis, we show in Section 3 spectrograms formed from BNL data and from Super-Kamiokande[9] data. In Section 4, we discuss the possible physical context: the role of the Resonant-Spin-Flavor Precession (RSFP) mechanism[10,11], and experiments that may be able to detect the possible mechanical influence of neutrinos on radionuclides.

## 2. Possible influence of solar rotation and of solar oscillations

To reduce the influence of systematic effects, we have formed the ratio of the $^{32}$Si and $^{36}$Cl count rates in the BNL experiment[6]. A power-spectrum analysis of this ratio shows a peak at 1 year$^{-1}$ with power S = 25, a peak at 11.2 year$^{-1}$ with power S = 21, and another at about 13 year$^{-1}$ with power S = 17. We consider a search band of width 10 – 15 year$^{-1}$ as a possible synodic rate (as measured from Earth) for internal solar rotation. [The corresponding sidereal (absolute) rotation band would be 11 – 16 year$^{-1}$.] By using the shuffle test[12] to obtain a robust estimate of the significance of the peak at 11.2 year$^{-1}$, we find that there is only one chance in 10$^6$ of finding this big a peak in that band by chance.

We have also carried out a power-spectrum analysis of the $^{226}$Ra data acquired by the PTB experiment.[7] This analysis also yields a peak close to 11.2 year$^{-1}$. In order to estimate the significance of the apparent rotational modulation in both the BNL and the PTB data, we have smoothed the two power spectra (which were oversampled) and then formed the joint power statistic.[13] We find that this statistic has the value 10.7 at 11.2 year$^{-1}$. By shuffling the BNL and PTB data separately, we find that there is only one chance in 10$^{17}$ of finding this big a value by chance. Hence the evidence for solar rotational modulation of nuclear decay rates is very strong.

Apart from the periodicity due to solar rotation, the strongest periodicity in solar data is the so-called "Rieger" oscillation, discovered by Rieger and his colleagues in their analysis of gamma-ray-flare data, which has a period of 154 days, corresponding to a frequency of 2.37 year$^{-1}$.[14] We have shown that this oscillation, which shows up in a wide range of solar data, may be attributed to an r-mode oscillation, [15] with spherical harmonic indices l = 3, m = 1, in the solar tachocline.[16] Detailed study of the evidence for rotational modulation of nuclear decay rates leads us to suspect that this modulation occurs in a region of the Sun where the sidereal rotation rate is in the band 12.0 – 13.7 year$^{-1}$. If a Rieger-type oscillation were to occur in that region, it would have a frequency in the range 2.00 to 2.28 year$^{-1}$. We have searched the combined BNL and PTB data for an oscillation in this band,[8] and found one at 2.11 year$^{-1}$, with a joint power statistic of 31. According to the shuffle test, there is only one chance in 10$^{12}$ of finding such a strong feature in that search band by chance.

### 3. Comparison with Super-Kamiokande solar-neutrino data

It appears that some form of radiation from the Sun influences certain nuclear decay rates. This radiation is influenced by one or more processes in the deep solar interior, but the radiation escapes from the Sun without losing that modulation. These requirements are suggestive of neutrinos: neutrinos can travel freely through the solar interior but, on the other hand, neutrinos can change flavor, either by the MSW (Mikheyev, Smirnov, Wolfenstein)[17,18] effect, or by the RSFP (Resonant Spin Flavor precession)[19,20] effect. Since the RSFP process involves magnetic field, and since the Sun's internal magnetic field is expected to be inhomogeneous and probably variable, it is possible that the emerging neutrino flux will have a structure and composition that has been influenced by the Sun's internal magnetic field. This could explain the rotational and r-mode modulations detected in decay data, since the internal magnetic field is influenced both by rotation and by these oscillations.

This leads us to examine the possibility that the solar neutrino flux, as detected by the Super-Kamiokande experiment, may exhibit oscillations similar to those detected in decay data. We have shown elsewhere that Super-Kamiokande measurements exhibit a periodicity with frequency 9.4 year$^{-1}$ [21] and a group of oscillations that may be attributed to r-mode oscillations in the tachocline,[22] but neither of these patterns is duplicated in decay data.

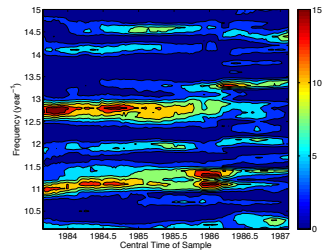

Fig 1. Spectrogram formed from BNL $^{36}$Cl data.

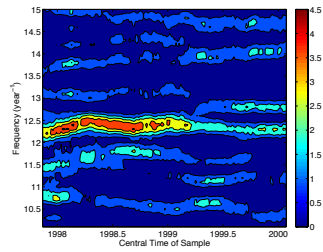

Fig 2. Spectrogram formed from Super-Kamiokande solar-neutrino data.

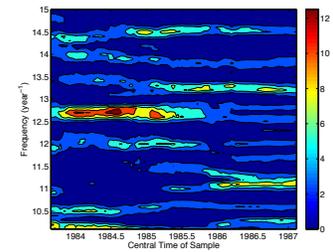

Fig 3. Spectrogram formed from BNL $^{32}$Si data.

However, there is a peak in the rotational search band, at frequency 12.7 year$^{-1}$ with power S = 5, and we have found that the power spectrum formed from the ratio of the $^{32}$Si and $^{36}$Cl BNL data contains significant peaks at 12.4 and 12.9 year$^{-1}$. Oscillations in decay data are typically transient and may drift in frequency. For this reason, and since there is no time overlap between the BNL and Super-Kamiokande datasets, we should not assume that an oscillation in one dataset would show up at the same frequency in the other dataset. To allow for possible intermittency and for drifting in frequency, we examine oscillations in BNL and Super-Kamiokande data by means of time-frequency analysis.

The above figures show spectrograms formed from BNL $^{36}$Cl and $^{32}$Si data (Figures 1 and 3, respectively) and a spectrogram formed from Super-Kamiokande data (Figure 2). We see that there is a prominent feature at or close to 12.5 year$^{-1}$ in all three figures.

### 4. Nuclear-physics processes

The results shown in the above figures suggest that neutrinos from the solar core are influenced by structures or processes in the solar interior. We now examine the possibility that this influence may be attributed to the RSFP effect.

The RSFP process occurs where two conditions are met [19,20]. One is the resonance condition

$$G_F\sqrt{2}(N_e - N_n) = \frac{\Delta(m^2)}{2E}, \text{ where } G_F = 10^{-37.03} \, eV \, cm^3,$$

and $N_e$, $N_n$ are the electron and neutron densities and E is the neutrino energy. The other is the adiabaticity condition, involving the scale height H, which we may write as

$$H > \frac{G_F\sqrt{2}(N_e - N_n)}{4(\mu/\mu_B)^2 \mu_B^2 B^2}, \text{ where } \mu_B = 10^{-7.23} \, eV \, G^{-1}.$$

It is convenient to examine these conditions in relation to the r-mode oscillations detected in Super-Kamiokande data[22]. These appear to be associated with the tachocline, which is a narrow region which we take to be located in the normalized radius range 0.68 to 0.72. At this location, the scale height is $6 \, 10^8$ cm, and $N_e - N_n$ is in the range $3 - 4 \, 10^{22}$ cm$^{-3}$. [23] If we assume that the neutrinos detected by Super-Kamiokande had energies in the range 5 – 6 MeV, we may infer from the resonance condition that $\Delta(m^2)$ is in the range $4 - 6 \, 10^{-8}$ eV$^2$, which is close to but less than the estimates ($1.3 \, 10^{-7}$ to $2.7 \, 10^{-6}$ eV$^2$) derived by Das et al.[24] from their global models. Adopting the estimate by Weber et al.[25] of the magnetic field strength in the tachocline (80 kG), we obtain the estimate $\mu/\mu_B > 3 \, 10^{-10}$, approximately. If the magnetic field strength is in fact higher by a factor of 3, $\mu/\mu_B$ would be reduced to $10^{-10}$.

In principle, it may be possible to measure the mechanical influence of the presumed neutrino flux on a nuclide. An energy transfer of $\Delta E_{eV}$ leads to a momentum transfer of $10^{-22.3} \Delta E_{eV}$ (g cm s$^{-1}$). If M (g) is the mass of the sample, A the atomic weight, $T_{yr}$ the half life in years, and if $f$ is the fraction of the decay rate that is due to the Sun, then we find that the force (in dynes) is given by

$$F = 10^{-6.2} f \, A^{-1} T_{yr}^{-1} \Delta E_{eV} M.$$

As an example, consider $^{54}$Mn, for which A = 54.9 and $T_y$ = 0.81. For the typical value f = 0.01 [26], we obtain

$$F \approx 10^{-10} \Delta E_{eV} M.$$

If, for instance, the energy transfer is 1 MeV and the mass is 1 milligram, we find that the force would be $10^{-7}$ dyne.

We may also consider the influence of spin transfer on a specimen. This leads to a mean torque per active atom of

$$H_1 = f\gamma\frac{1}{2}h = 10^{-34.1} f \, T_{yr}^{-1},$$

where $\gamma$ is the decay coefficient in s$^{-1}$. We consider a vertical cylinder of radius R suspended freely. Then if $K$ is the fraction of the mass due to the nuclide, and if $\Theta$ is the maximum solar elevation, we find that the amplitude of the daily oscillation (in radians) is given by

$$\phi_0 = 10^{-1.8} f \, K \, A^{-1} T_{yr}^{-1} R^{-2} \sin(\Theta).$$

For $^{54}$Mn and for $\Theta$ = 45 deg, we obtain $\phi_0 = 10^{-3.6} K R^{-2}$. For K = 0.01 and R = 0.01 cm, we obtain an amplitude of 0.025 radian, or about 1 degree.

The work of EF was supported in part by US DOE contract No. DE-AC02-76ER071428.